\documentclass[reqno]{amsproc}

\usepackage{amstext,amsmath,amssymb,amsfonts}
\usepackage[latin1]{inputenc}
\usepackage{epsfig}
\usepackage{hyperref}
\usepackage{color}
\usepackage{amssymb}
\usepackage{amsfonts}
\usepackage{amscd}
\usepackage{xcolor}

\usepackage{latexsym}

\newtheorem{definition}{Definition}
\newtheorem{theorem}{Theorem}

\newcommand{\1}{1 \! \! {\rm I}}

\newcommand{\bea}{\begin{eqnarray}}
\newcommand{\eea}{\end{eqnarray}}
\newcommand{\beq}{\begin{equation}}
\newcommand{\eeq}{\end{equation}}
\newcommand{\enn}{\nonumber \end{equation}}

\title[Control of ODE using Bagarello's approach]{Control of ordinary differential equations 
using Bagarello's operator approach :  
Case of forced harmonic oscillator systems\footnote{Preprint: ICMPA-MPA/2015/09 }
}

\author{Jean Ghislain Compaor\'e}
\address[J.G.C.]{International Chair in Mathematical Physics and Applications,
ICMPA-UNESCO Chair, 072 BP 50, Cotonou, Rep. of Benin}
\email{ghislaincompaore@yahoo.fr}

\author{Villevo Adanhounm\`e}
\address[V.A.]{International Chair in Mathematical Physics and Applications,
ICMPA-UNESCO Chair, 072 BP 50, Cotonou, Rep. of Benin}
\email{adanhounmvillvo@yahoo.fr}%

\author{Mahouton Norbert Hounkonnou}
\address[M.N.H.]{International Chair in Mathematical Physics and Applications,
ICMPA-UNESCO Chair, 072 BP 50, Cotonou, Rep. of Benin}
\email{norbert.hounkonnou@cipma.uac.bj}

\begin{document}

\maketitle 
\today
\begin{abstract}
This work deals with the study of an optimal control of a system of nonlinear 
differential equations  using the Bagarello's operator approach recently 
introduced in a paper ({\it Int. Jour. of Theoretical Physics}, 
\textbf{43}, issue 12 (2004), p. 2371 - 2394). 
The control problem is reduced, by using the Pontryagin's maximum principle, 
to a system of ordinary differential equations with unknown 
state and adjoint variables. 
Its solution  is then described in terms of  a series expansion of commutators involving an unbounded
self-adjoint,   densely defined, system Hamiltonian operator $H$ and initial position operators.
Relevant simple applications  are discussed.

\end{abstract}
{\noindent
 

\tableofcontents

\section{Introduction}\label{intro}
Various methods are used to solve systems of ordinary differential equations (SODEs) 
in mathematics and applied sciences. The most popular ones include factorization 
\cite{Hounkonnou}, linearization \cite{Krasnov}, perturbation \cite{Holmes,Nayfeh}, 
closure approximation \cite{Adomian,Adomian1}, discretization \cite{Stetter} and
Adomian methods \cite{Adomian,Adomian2,Cherruault,Cherruault1,Sielenou} to cite a few.
Recently, in an interesting paper \cite{Bagarello97}, Bagarello 
  developed a non-commutative method, based on the quantum mechanics formalism, for the  analysis of  
systems of ODEs. He provided  solutions of some systems described by  unbounded 
self-adjoint and densely defined Hamiltonian operators and 
discussed  corresponding integrals of motion.

The possibility of acting, in an appropriate way,  on systems 
of ODEs governing physical phenomena or other mechanisms in nature
gives rise to  the control theory.  
  Optimal control problems 
can be solved by the so-called Pontryagin's method.

One of the most important systems encountered in the literature is certainly 
the system of harmonic oscillators, widely used  as a basic tool in physics 
\cite{Barone,MoshinskyAl}. This is motivated by their role in many applications 
in various fields of physics and technology. The harmonic oscillators describe a wide class of physical models. Besides, their properties are well known. This explains why they 
are often used as a preliminary tool in order to gain initial  insight into  complex 
systems. For instance, quantum mechanics  as well as optimal control were first 
illustrated using systems of harmonic oscillators. A typical controlled harmonic 
oscillator is a pendulum made of a string and a ball moving in a vertical plane 
and subjected to a force (also called the control) whose role is to stress the system to 
rest in a minimum time.

Recently, Andresen et al \cite{Salamon} invesigated the control of an oscillator using 
the frequency variation.
Dasanayake \cite{Dasanayake} and Van Dooren \cite{VanDooren}, 
(see also references therein), performed numerical solutions of optimal control problems 
 by  pseudospectral methods and Chebyshev series.

In the present work,  for the first time to the best  of our knowledge,  the  Bagarello's noncommutative approach is applied to the control of systems of nonlinear ordinary differential equations.

The paper is organized as follows. In Section 2, we present  the considered optimal control problem. We derive the state and adjoint 
systems of ordinary differential equations  using the Pontryagin's maximum 
principle \cite{Gamkrelidze,PontryaginAl}.
Then, we use  the Bagarello's operator method to solve such  systems. 
In Section 3,   harmonic oscillator  systems are controled and
 discussed. 

\section{Theoretical framework}\label{sec:1}
We consider the following optimal control problem \cite{Arutyunov,Tynyanskii}:
\begin{eqnarray} 	\dot{x} = f(x, u), & \quad x(0) = x_0 \label{CO01} \\ x(T) = x_1 &  \\
 	u \in V \subset E^r & \\ \min_{u \in V} \int_0^T f^0 (x,u) dt, & 
\end{eqnarray}
where $x, x_0, x_1 \in E^n;$
 $E^r\supset V \ni u$ is the control, $V$ the control set; $T$ 
stands for the moment and  $E$ is the Euclidean space. 
All the functions occurring in the formulation of 
the problem are assumed to be differentiable with respect to $(x, u).$ 
Further, the minimum should be in a class of measurable and bounded 
functions $u,$ defined on $\left[0, T\right].$ 
For instance, $V$ can be represented in the form 
\[ V = \left\{u \in E^r : h_i(u) \leq 0, i\in I_1; h_i(u) = 0, i\in I_2 \right\} \] 
where the functions $h_i$   express the constraints on the control; 
 $I_1$ and $I_2$ are finite disjoint index sets.

Let us introduce the Hamilton-Pontryagin's function 
\[ \eta(x, u, \psi) = \psi f^*(x,u) - \psi^0 f^0(x,u) \]
where $ f = (f_1, \dots, f_n) \in E^n; \quad E^n \ni \psi = (\psi_1, \dots, \psi_n) $ is
the adjoint variable depending on $t; *$ stands for the transposition operation.  $\psi^0 $ is a positive number.

By virtue of the Pontryagin's maximum principle \cite{Gamkrelidze,PontryaginAl}, 
 the existence of adjoint functions 
is a necessary condition to link the control system of ordinary differential 
equations to a functional objective. The 
 optimal control $u^*$ is then characterized 
  in terms 
of the state and adjoint functions.

Thus, given an optimal control $u^*$ and the corresponding system (\ref{CO01}), 
there exists an adjoint variable $\psi$ satisfying the following equation: 
\begin{equation}\label{eq03}\frac{d \psi}{dt} = - \frac{\partial \eta}{\partial x} 
(x, u^*(x, \psi), \psi). \end{equation}
There results  the Cauchy problem 
\begin{eqnarray} \dot{x} &=& f(x, u^*(x, \psi)), \quad x(0) = x_0 \\
	\dot{\psi} &=& - \frac{\partial \eta}{\partial x} 
	(x, u^*(x, \psi), \psi), \quad \psi (0) = \psi_0  
\end{eqnarray} with the unknowns $x, \psi.$

Setting $y = \begin{pmatrix} x \\ \psi \end{pmatrix},
g(y) = \begin{pmatrix} f(x, u^*(x, \psi)) \\ 
- \dfrac{\partial \eta}{\partial x} (x, u^*(x, \psi), \psi) \end{pmatrix}, 
y(0) =  \begin{pmatrix} x(0) \\ \psi(0) \end{pmatrix},$

we can write the latter in the form:  
\begin{equation}\label{SODE} \dot{y} = g(y), \quad y(0) = y_0 = 
\begin{pmatrix} x_0 \\ \psi_0 \end{pmatrix}, \end{equation}
or, explicitly,
\begin{equation}\label{SODE1} \left\{ \begin{array}{lcl}
	\dot{y}_1 &=& g_1(y_1, y_2, \dots, y_{2n}) \\ 
	\dot{y}_2 &=& g_2(y_1, y_2, \dots, y_{2n}) \\ \vdots &=& \vdots \\
	\dot{y}_{2n} &=& g_{2n}(y_1, y_2, \dots, y_{2n}) \\ 
	&& y_j(0) = y_j^0, \quad j = 1, \dots, 2n. \end{array} \right. \end{equation}
We suppose that the functions $g_j$  are such that the solution 
of the Cauchy problem (\ref{SODE1}) exists and is unique. 

In this work, we intend to solve this system by Bagarello's approach (\cite{Bagarello97}, 
and references therein), based on a  quantum mechanical formalism. 
Given a quantum mechanical system $\mathcal{S}$ and 
the related set of observables $\mathcal{O}_\mathcal{S},$ i. e.,
the set of all  self-adjoint bounded (or more often unbounded)
operators describing $\mathcal{S},$ the evolution of any observable 
$Y \in \mathcal{O}_\mathcal{S}$ satisfies the Heisenberg 
equation of motion (HOEM)\cite{Bagarello97}:
		\begin{equation}\label{eqNC01} \frac{d}{dt} Y(t) = i[H, Y(t)]. \end{equation}
Here $[A, B] = AB - BA $ is the commutator between $A, B \in \mathcal{O}_\mathcal{S}; 
H$ is assumed to be a densely defined self-adjoint Hamiltonian operator of the system   
acting on some Hilbert space $\mathcal{H},$ given by \cite{Bagarello97}
			 \begin{equation}\label{eqNC02} H(\vec{g}_0) =  \frac{1}{2} \sum_{j=1}^{2n} 
	\left\{ p_j g_j(y_1^0, y_2^0, \dots, y_{2n}^0) 
	+ g_j(y_1^0, y_2^0, \dots, y_{2n}^0) p_j \right\} \end{equation}
where $\vec{g}_0 = \left(g_1(\vec{Y}^0), g_2(\vec{Y}^0), \dots, g_{2n} (\vec{Y}^0) \right),
  \vec{Y}^0 = (y_1^0, y_2^0, \dots, y_{2n}^0) ; $ 
the initial position $y_j^0$ is considered as an operator acting 
on the Hilbert space $\mathcal{H}$, and $p_j$ is its canonical conjugate  momentum operator
such that 
\begin{eqnarray} \left[y_j^0, p_k \right] & = & 
i \delta_{j,k} \1, \quad  j, k = 1, \dots, 2n\\ 
\left[y_j^0, y_k^0\right] & = & \left[p_j, p_k\right] = 0, 
\quad j, k = 1, \dots, 2n. \end{eqnarray}
Standard results in quantum mechanics show that for any differentiable
functions \[ \varphi\left(y_1^0, y_2^0, \dots, y_{2n}^0\right),  \text{ and  } 
\hat{\varphi}\left(p_1, p_2, \dots, p_{2n}\right), \text{ we have }\]
				\begin{eqnarray} \left[p_j, \varphi(y_1^0, y_2^0, \dots, y_{2n}^0)\right] & = &
				- i \frac{\partial \varphi}{\partial y_j^0}, \quad j = 1, \dots, 2n \label{eqNC07a} \\ 
				\left[y_j^0, \hat{\varphi}\left(p_1, p_2, \dots, p_{2n}\right) \right] & = & 
				i \frac{\partial \hat{\varphi}}{\partial p_j}, \quad j = 1, \dots, 2n. \label{eqNC07b}
				\end{eqnarray}
\begin{theorem}\cite{Bagarello97} If the functions $g_j$ are holomorphic, 
a  formal solution of the HOEM (\ref{SODE1}) is 
\begin{equation}\label{eqNC02a} Y(t) = e^{iHt} Y^0 e^{-iHt}.\end{equation}
where $Y^0$	is the initial value of $Y(t)$ and $H$ does not depend explicitly on time.
			 	
Furthermore if $H$ is bounded, we get 
\begin{equation}\label{eqNC04} Y(t) = \sum_{k \geq 0}  
\frac{(it)^k}{k!} [H, Y^0]_k \end{equation}
where $[A, B]_k $ is the multiple commutator recursively defined  as :
\[ [A, B]_0 = B ; \quad  [A, B]_k = [A, [A, B]_{k-1}].\]  	\end{theorem}

The concepts of integral of motion and extended integral of motion 
of the system (\ref{SODE1}) are defined as in \cite{Bagarello97}.
\begin{definition} Any holomorphic function $I$ depending on the variables $y_j,$ such that 
\[I(y_1(t), y_2(t), \dots, y_{2n}(t)) = I_0 
\quad \forall t, \quad I_0 = \text{ const., }\]
is called an Integral of motion (IoM) of system (\ref{SODE1}).\end{definition}
\begin{definition} We call extended integral of motion (EIoM) of the system (\ref{SODE1})
any holomorphic function $J$ depending on the $y_j, p_j $ such that
\[ J(y_1(t), y_2(t), \dots, y_{2n}(t), p_1(t), p_2(t), \dots, p_{2n}(t) ) 
= J_0 \quad \forall t, J_0 = \text{ const.,} \]  
where $ y_j(t) = e^{iHt}y_j^0 e^{-iHt} $ and $p_j(t) 
= e^{iHt} p_j e^{-iHt}, j = 1, \dots, 2n.$
\end{definition}	      
				
One of the main advantages of the Bagarello's strategy is to provide at hand 
a good approximation of the solution of the SODE.
The $N-$th order approximation ($N \in \mathbb{N}$) is 
\begin{equation}\label{approx} Y_N(t) = \sum_{k =0}^N  
\frac{(it)^k}{k!} [H, Y^0]_k. \end{equation}
Using the integral of motion, we can estimate the approximation's error 
of the unknown exact solution of the derived Cauchy problem.
In this case, one can proceed as follows : let $Y_N(t)$ be the 
approximated solution of $Y(t)$. We may thus compute and evaluate
the error \cite{Bagarello97}                                       
\begin{equation}\label{approxError} \Delta_N(t) = I(Y_N(t)) - I(Y^0). \end{equation} 

This approach is useful for our study because formula (\ref{approx})
provides a simple approximation scheme for exact solutions to ordinary
differential equations. Further the estimation of the error $\Delta_N(t)
$
tells us how good this approximation is.

In the next section, we apply this approach 
to time-optimal control problems.

\section{Applications}\label{sec:2}
\subsection{Problem 1}\label{sec:3} 
In classical mechanics, we study the forced harmonic oscillator. 
The corresponding optimal control problem has the form \cite{LandauI} 
\begin{equation} \label{prob101} \left\{ \begin{array}{lcl} \dot{x}_1 &=& x_2 \\ 
\dot{x}_2 &=& - \omega^2 x_1 + u(t) \\ x_1(0) &=& x_1^0 ; \quad  x_2(0) = x_2^0 ; \\
x_1(T) &=& 0  ; \quad x_2(T) = 0 ;\\ -1 &\leq& u \leq 1 \\ T &&\longrightarrow \inf.  
\end{array} \right. \end{equation}
Here $t$ and $T$ denote the time, $x = (x_1, x_2) \in E^2,$   
$u$ is the control and $\omega$ the oscillation frequency.

In this problem, the infimum is sought in a class of controls $u(t), t \geq 0.$ 
We study a control $u$ that moves the point $x(0)$ to the point $x(T) = 0$ 
in accordance with the corresponding solution of (\ref{prob101}) during the 
time $T.$ Then we solve the boundary value problem 
derived from the Pontryagin's maximum principle.

Let us consider the Hamilton-Pontryagin's function 
\begin{equation}\label{prob107}	\eta (x, u, \psi) = 
- 1 + \psi_1 x_2 + \psi_2 (- \omega^2 x_1 + u) \end{equation} 
where $\psi = (\psi_1, \psi_2) \in E^2.$

According to Pontryagin's maximum principle, we obtain the control $u$ 
in the form\[ u(t) = \left\{\begin{array}{lc}  1 & \text{  if  } 
\psi_2 (t) > 0 \\ - 1 & \text{  if  } \psi_2 (t) < 0 .  \end{array}\right..\]
Now we solve the Cauchy problem derived from the Pontryagin's 
maximum principle using the technique developed by Bagarello.\\
In the case $u(t)=1,$ we can write 
\begin{equation}\label{prb109}\left\{\begin{array}{l} 
\dot{x}_1 = x_2 \\ \dot{x}_2 = - \omega^2 x_1 + 1 \\ 
\dot{\psi}_1 = \omega^2 \psi_2 \\ \dot{\psi}_2 =
 - \psi_1 \end{array}\right.\end{equation}
\[x_1(0) = x_1^0 ; \quad x_2(0) = x_2^0 ; \quad 
\psi_1(0) = \psi_1^0 ; \quad \psi_2(0) = \psi_2^0. \]
Introducing the new variable $\tilde{x}_1 = x_1 - \dfrac{1}{\omega^2}, $ we rewrite (\ref{prb109})
in the form 
\begin{equation}\label{prb113}\left\{\begin{array}{l} 
\dot{\tilde{x}}_1 = x_2 \\ \dot{x}_2 = - \omega^2 \tilde{x}_1 \\ 
\dot{\psi}_1 = \omega^2 \psi_2 \\ \dot{\psi}_2 = 
- \psi_1 \end{array}\right.\end{equation}
In the case $u(t) = - 1$  we get 
\begin{equation}\label{prb115}\left\{\begin{array}{l} 
\dot{x}_1 = x_2 \\ \dot{x}_2 = - \omega^2 x_1 - 1 \\ 
\dot{\psi}_1 = \omega^2 \psi_2 \\ \dot{\psi}_2 = 
- \psi_1 \end{array}\right.\end{equation}
or, equivalently, denoting $\hat{x}_1 = x_1 + \dfrac{1}{\omega^2},$
 \begin{equation}\label{prb117}\left\{\begin{array}{l} 
\dot{\hat{x}}_1 = x_2 \\ \dot{x}_2 = - \omega^2 \hat{x}_1 \\ 
\dot{\psi}_1 = \omega^2 \psi_2 \\ \dot{\psi}_2 = 
- \psi_1 \end{array}\right.\end{equation}
with the previous initial conditions .

Equations (\ref{prb113}) and (\ref{prb117}) reduce to the equivalent system 
\begin{equation}\label{prb119}\left\{\begin{array}{l} \dot{y}_1 = y_2 \\ \dot{y}_2 = - \omega^2 y_1 \\ 
\dot{y}_3 = \omega^2 y_4 \\ \dot{y}_4 = - y_3 \end{array}\right.\end{equation} 
where $y_1 = \tilde{x}_1$ or $y_1 = \hat{x}_1, y_2 = x_2, y_3 = \psi_1, y_4 = \psi_2.$

Here, instead of using 
standard methods widely spread in the literature,  we  
 show that Bagarello's suggestion  can also be exploited to solve this optimal control problem and to reproduce known results. Indeed,
using the definition of the commutator
\[ g_j p_j = - \left[p_j, g_j \right] + p_j g_j, \quad j = 1, \dots, 4, \]
we compute the Hamiltonian of the previous system in the form
\begin{equation} H = p_1 y_2^0 - \omega^2 p_2 y_1^0 +  \omega^2 p_3 y_4^0 - p_4 y_3^0. \end{equation}
Developing the solution of (\ref{prb119}) as an infinite series:
\begin{eqnarray*} \begin{pmatrix} y_1(t) \\ y_2(t) \\ y_3 (t) \\ y_4(t) \end{pmatrix}
					  & = & \begin{pmatrix} y_1^0 \\y_2^0 \\y_3^0 \\y_4^0 \end{pmatrix} 
					        + it \left[H, \begin{pmatrix} y_1^0 \\y_2^0 \\y_3^0 \\y_4^0 \end{pmatrix} \right] 
					        + \frac{(it)^2}{2!} \left[H, \begin{pmatrix} y_1^0 \\
					        y_2^0 \\y_3^0 \\y_4^0 \end{pmatrix} \right]_2 
					        + \frac{(it)^3}{3!} \left[H, \begin{pmatrix} y_1^0 \\
					        y_2^0 \\y_3^0 \\y_4^0 \end{pmatrix} \right]_3 + \dots				\end{eqnarray*} 
we calculate the multiple commutator 
\[ \begin{pmatrix} \left[ H,  y_1^0 \right]_{2p} \\ \left[H,  y_2^0 \right]_{2p} \\
					      \left[ H,  y_3^0 \right]_{2p} \\ \left[H,  y_4^0 \right]_{2p} \end{pmatrix} = 
					      \begin{pmatrix} \omega^{2p}  y_1^0 \\ \omega^{2p}  y_2^0 \\ 
					      \omega^{2p}  y_3^0 \\ \omega^{2p}  y_4^0 \end{pmatrix}  ; \quad 
					       \begin{pmatrix} \left[H,  y_1^0 \right]_{2p+1} \\ \left[H,  y_2^0 \right]_{2p+1} \\
					      \left[H,  y_3^0 \right]_{2p+1} \\ \left[H,  y_4^0 \right]_{2p+1} \end{pmatrix} = 
					      \begin{pmatrix} - i\omega^{2p}  y_2^0 \\ i \omega^{2p+2}  y_1^0 \\ 
					      - i\omega^{2p+2}  y_4^0 \\ i \omega^{2p}  y_3^0 \end{pmatrix},  p = 0, 1, 2, 3, \dots \]
and obtain			
				\begin{eqnarray*} \begin{pmatrix} y_1(t)\\ y_2(t) \\y_3(t)\\ y_4(t) \end{pmatrix} 
				& = & \begin{pmatrix}  y_1^0 \cos (\omega t) + \dfrac{ y_2^0}{\omega} \sin (\omega t) \\ 
				 y_2^0 \cos (\omega t) -  y_1^0 \omega \sin (\omega t) \\ 
				  y_3^0 \cos (\omega t) + y_4^0 \omega \sin (\omega t) \\ 
				 y_4^0 \cos (\omega t) -  \frac{ y_3^0}{\omega} \sin (\omega t)\end{pmatrix}. \end{eqnarray*}
In order to determine the time $T$ such that 
\[ \left\{\begin{array}{l} x_1 (T) = 0 \\ x_2  (T) = 0 \end{array}\right., \]
and taking into account $y_1 = \tilde{x}_1 = x_1 - \dfrac{1}{\omega^2} , y_2 = x_2,$ 
we obtain
			\[ \begin{pmatrix} x_1(t) \\ x_2(t) \end{pmatrix} = 
			\begin{pmatrix} (x_1^0 - \frac{1}{\omega^2})\cos (\omega t) + 
			\frac{x_2^0}{\omega} \sin (\omega t) + \frac{1}{\omega^2}  \\
			x_2^0 \cos (\omega t) - (x_1^0 - \frac{1}{\omega^2}) \omega \sin (\omega t)  \end{pmatrix} \]
			and  \[ \cos (\omega T) = \frac{\omega}{x_2^0} \Big(x_1^0 - \frac{1}{\omega^2}\Big) \sin (\omega T) ; \quad
			\sin (\omega T) = \frac{-x_2^0 / \omega}{ \omega^2 (x_1^0 - \frac{1}{\omega^2})^2 + (x_2^0)^2}.\] 

Thus  \[ \tan \omega T  = \frac{x_2^0}{\omega (x_1^0 - \frac{1}{\omega^2})} \quad \text{  or } \quad
			\omega T = \arctan\frac{x_2^0}{\omega (x_1^0 - \frac{1}{\omega^2})} ; \]
			\[ \tilde{T} = \frac{1}{\omega} \arctan\frac{x_2^0}{\omega (x_1^0 - \dfrac{1}{\omega^2})}, 
			\quad x_2^0 \left(x_1^0 - \frac{1}{\omega^2}\right) > 0 \]

Analogously, for $y_1 = \hat{x}_1 = x_1 + \dfrac{1}{\omega^2} , y_2 = x_2,$ we get 
\[ \begin{pmatrix} x_1(t) \\ x_2(t) \end{pmatrix} 
=  \begin{pmatrix} \left(x_1^0 + \frac{1}{\omega^2}\right)\cos (\omega t) 
+ \frac{x_2^0}{\omega} \sin (\omega t) - \frac{1}{\omega^2} \\
x_2^0 \cos (\omega t) - \left(x_1^0 + \frac{1}{\omega^2}\right) \omega \sin (\omega t) \end{pmatrix}\]
and the relations \[ \cos (\omega T) = \frac{\omega}{x_2^0} (x_1^0 + \frac{1}{\omega^2}) \sin (\omega T) ; \quad
\sin (\omega T) = \frac{x_2^0 / \omega}{ \omega^2 (x_1^0 + \frac{1}{\omega^2})^2 + (x_2^0)^2}\] 
implying \[ \tan \omega T  = \frac{x_2^0}{\omega (x_1^0 + \frac{1}{\omega^2})} \quad \text{  or } \quad
\omega T = \arctan\frac{x_2^0}{\omega (x_1^0 + \frac{1}{\omega^2})} ; \]
\[ \hat{T} = \frac{1}{\omega} \arctan\frac{x_2^0}{\omega (x_1^0 + \frac{1}{\omega^2})}, 
\quad x_2^0 \left(x_1^0 + \frac{1}{\omega^2}\right) > 0. \]
Therefore, the optimal time solution to problem (\ref{prob101}) 
is given by considering the following  two cases :
if $x_2^0 > 0,$ then $\hat{T}$ is the optimal time 
and if $x_2^0 < 0,$ then $\tilde{T}$ is the optimal time.  

In the sequel, 
we 
deal  with more complicated problems whose solutions can be obtained 
 by perturbative approaches. In this case, we   show that as a 
possible	 candidate, Bagarello's formalism can be easily implemented
in a suitable and solvable form.				      
\subsection{Problem 2}\label{sec:4}
Let us consider the optimal control problem of a pendulum with large oscillations in  the form \cite{Vassiliev}
\begin{equation} \label{prob201} \left\{ \begin{array}{lcl} 
 \dot{x}_1 &=& x_2 \\ \dot{x}_2 &=& - \beta x_2 - \sin x_1 + u(t) \\ 
x_1(0) &=& x_1^0 ; \quad  x_2(0) = x_2^0 ; \\x_1(T) &=& 0  ; \quad x_2(T) = 0 ; \\ 
-1 &\leq& u \leq 1 \\ T & \longrightarrow & \inf \end{array} \right. \end{equation}
 $ u(t) = \dfrac{1}{m} F(t), \; 0 \leq t \leq T,  $ 
is the control subjected to the following constraint : 
\begin{equation} u \in V =\left\{u \in E^1 : \left|u\right| \leq 1 \right\}; \end{equation}
$x_1^0, x_2^0$ are given positive constants, 
$m$ is the mass, $F$ the force and $\beta >0 $ a constant.

The Hamilton-Pontryagin's function is written as
\begin{equation}\label{prob207}
	\eta (x, u, \psi) = - 1 + \psi_1 x_2 + \psi_2 (- \beta x_2 - \sin x_1 + u) 
\end{equation} where $\psi = (\psi_1, \psi_2) \in E^2.$

By virtue of the Pontryagin's maximum principle, the control is given by  
\[ u(t) = \left\{\begin{array}{lc}  1 & \text{  if  } \psi_2 (t) > 0 \\ 
- 1 & \text{  if  } \psi_2 (t) < 0.  \end{array}\right.\]
The associated Cauchy problem 
reads
\begin{equation}\label{prb209}\left\{\begin{array}{l} \dot{x}_1 = x_2 \\ 
\dot{x}_2 = - \beta x_2 - \sin x_1 + u(t) \\ 
\dot{\psi}_1 = \psi_2 \cos x_1 \\ 
\dot{\psi}_2 = - \psi_1 + \beta \psi_2 \end{array}\right.\end{equation}
with the initial conditions
\[x_1(0) = x_1^0, x_1^0 > 0  ; \quad x_2(0) = x_2^0, x_2^0 > 0 ; \quad 
\psi_1(0) = \psi_1^0 ; \quad \psi_2(0) = \psi_2^0. \]
Equivalently, this can be re-expressed as follows:  
\begin{equation}\label{prb211}\left\{\begin{array}{l} \dot{y}_1 = y_2 \\ 
\dot{y}_2 = - \beta y_2 - \sin y_1 + u(t) \\ 
\dot{y}_3 = y_4 \cos y_1 \\ \dot{y}_4 = - y_3 + 
\beta y_4 \end{array}\right.\end{equation}
\[ y_1(0) = y_1^0 ; \quad y_2(0) = y_2^0 ; 
\quad y_3(0) = y_3^0 ; \quad y_4(0) = y_4^0 \] 
where $y_1 = x_1, y_2 = x_2, y_3 = \psi_1, y_4 = \psi_2.$

The Hamiltonian of the system takes the form: 
\begin{equation}\label{prb213} H = p_1 y_2^0 + p_2 
\left[- \beta y_2^0 - \sin y_1^0 + u(t)\right] + p_3 y_4^0 \cos y_1^0 + p_4 (- y_3^0 + \beta y_4^0). \end{equation}

The solution  (\ref{approx}) becomes in this case: 
\begin{eqnarray*}	 Y_2 (t) & = & \begin{pmatrix} 
y_1^0 \\y_2^0 \\y_3^0 \\y_4^0 \end{pmatrix} 
+ it \left[H, \begin{pmatrix} y_1^0 \\y_2^0 \\
y_3^0 \\y_4^0 \end{pmatrix} \right] 
+ \frac{(it)^2}{2!} \left[H, \begin{pmatrix} 
y_1^0 \\y_2^0 \\y_3^0 \\y_4^0 \end{pmatrix} \right]_2 
\end{eqnarray*} where the commutators are given by \\
\[\begin{pmatrix} \left[ H,  y_1^0 \right] \\ \left[H,  y_2^0 \right] \\
\left[ H, y_3^0 \right]\\ \left[H,  y_4^0 \right] \end{pmatrix} = 
\begin{pmatrix} -i  y_2^0 \\ -i \left[-\beta y_2^0 - \sin y_1^0 + u(t)\right]\\ 					      
-i y_4^0 \cos y_1^0 \\ - i (-y_3^0 + \beta y_4^0) \end{pmatrix} ;\] 
					      
\[ \begin{pmatrix} \left[H, y_1^0 \right]_2 \\ \left[H, y_2^0 \right]_2 \\
\left[H, y_3^0 \right]_2 \\ \left[H, y_4^0 \right]_2 \end{pmatrix} = 
\begin{pmatrix} (- i)^2 \left[- \beta y_2^0 - \sin y_1^0 + u(t)\right] \\ 
(-i)^2 \left\{- \beta \left[- \beta y_2^0 - \sin y_1^0 + 
u(t)\right] - y_2^0 \cos(y_1^0) \right\} \\ 
(-i)^2 \left[-  y_2^0 y_4^0 \sin(y_1^0) + 
(-y_3^0 + \beta y_4^0 )\cos(y_1^0)\right]\\ 
(-i)^2 \left[- y_4^0 \cos(y_1^0) + \beta 
(-y_3^0 + \beta y_4^0)\right] \end{pmatrix}.\]			      

Finally we get \begin{itemize}
\item For $u(t) = - 1$, \[ \begin{pmatrix} 
\tilde{y}_{1.2}(t) \\ \tilde{y}_{2.2}(t) \end{pmatrix} 
= \begin{pmatrix} y_1^0 + t y_2^0 + 
\frac{t^2}{2}(-\beta y_2^0 - \sin y_1^0 - 1)\\
y_2^0 + t (-\beta y_2^0 - \sin y_1^0 - 1)
+ \frac{t^2}{2} \left[- 
\beta (- \beta y_2^0 - \sin y_1^0 - 1) 
- y_2^0\cos y_1^0 \right] \end{pmatrix} \]
and 
\[ \begin{pmatrix} \tilde{x}_{1.2}(t) \\ \tilde{x}_{2.2}(t) \end{pmatrix} 
= \begin{pmatrix} x_1^0 + t x_2^0 + \frac{t^2}{2}(-\beta x_2^0 - \sin x_1^0 - 1)\\
x_2^0 + t (- \beta x_2^0 - \sin x_1^0 -1) + \frac{t^2}{2} \left[- \beta (- \beta x_2^0 
- \sin x_1^0 - 1) - x_2^0 \cos x_1^0  \right] \end{pmatrix} \]
											
\item For $u(t) = 1$, \[ \begin{pmatrix} \hat{y}_{1.2}(t) \\ 
\hat{y}_{2.2}(t) \end{pmatrix}
= \begin{pmatrix} y_1^0 + t y_2^0 + \frac{t^2}{2}
(-\beta y_2^0 - \sin y_1^0 + 1)\\
y_2^0 + t (-\beta y_2^0 - \sin y_1^0 + 1)
+ \frac{t^2}{2} \left[- \beta 
(- \beta y_2^0 - \sin y_1^0 + 1) 
- y_2^0\cos y_1^0 \right] \end{pmatrix} \] and
\[ \begin{pmatrix} \hat{x}_{1.2}(t) \\ 
\hat{x}_{2.2}(t) \end{pmatrix}
= \begin{pmatrix} x_1^0 + t x_2^0	+ 
\frac{t^2}{2}(-\beta x_2^0 - \sin x_1^0 + 1)\\
x_2^0 + t (- \beta x_2^0 - \sin x_1^0 + 1) + 
\frac{t^2}{2} \left[- \beta (- \beta x_2^0 
- \sin x_1^0 + 1)	- x_2^0 \cos x_1^0  \right] \end{pmatrix} .\]
\end{itemize}
Taking into account the relations 
\[ \left\{\begin{array}{l} \tilde{x}_{1.2} (T) = 0 \\ 
\tilde{x}_{2.2}  (T) = 0 \end{array}\right. 
\text{  and  } \left\{\begin{array}{l} \hat{x}_{1.2} (T) 
= 0 \\ \hat{x}_{2.2}  (T) = 0 \end{array}\right., \]
we obtain 
\begin{itemize}	\item[(i)] $ \tilde{T} =\dfrac{(-\beta x_2^0 - \sin x_1^0 - 1) 
(x_2^0+\beta x_1^0) + x_1^0 x_2^0 \cos x_1^0 }
{(-\beta x_2^0 - \sin x_1^0 - 1)(\sin x_1^0+1) - (x_2^0)^2 \cos x_1^0}$ 
\[\text{and }  a_2 \left[(a_1 c_2 - a_2 c_1)^2 + 
(a_2 b_1 - a_1 b_2) (b_1 c_2 - b_2 c_1)\right]= 0 \] with  
$a_1, b_1, c_1$ and $a_2, b_2, c_2$ 
being respectively the coefficients of the  polynomials $\tilde{x}_{1.2}(t)$ and $\tilde{x}_{2.2}(t).$
\item[(ii)] $ \hat{T} =\dfrac{(-\beta x_2^0 - 
\sin x_1^0 + 1) (x_2^0+\beta x_1^0) + x_1^0 x_2^0 \cos x_1^0 }
{(-\beta x_2^0 - \sin x_1^0 + 1)(\sin x_1^0-1) - (x_2^0)^2 \cos x_1^0}$
\[ \text{and }  A_2 \left[(A_1 C_2 - A_2 C_1)^2 + 
(A_2 B_1 - A_1 B_2) (B_1 C_2 - B_2 C_1)\right]= 0 \] with 
$A_1, B_1, C_1$ and $A_2, B_2, C_2$ 
being respectively the coefficients of the  polynomials $\hat{x}_{1.2}(t)$ and $\hat{x}_{2.2}(t).$ 
\end{itemize}

If $ \dfrac{(-\beta x_2^0 - \sin x_1^0 - 1) 
(x_2^0+\beta x_1^0) + x_1^0 x_2^0 \cos x_1^0 }
{(-\beta x_2^0 - \sin x_1^0 - 1)(\sin x_1^0+1) - (x_2^0)^2 \cos x_1^0} < $ 
\[\frac{(-\beta x_2^0 - \sin x_1^0 + 1) 
(x_2^0+\beta x_1^0) + x_1^0 x_2^0 \cos x_1^0 } 
{(-\beta x_2^0 - \sin x_1^0 + 1)(\sin x_1^0-1) - (x_2^0)^2 \cos x_1^0} \]
then $\tilde{T}$ is the solution of the problem. 
Otherwise $\hat{T}$ is the solution.

Let us estimate the error using the following integral of motion  
\[ I(x_1, x_2) = x_2 + \beta x_1  + \int_0^t \left[\sin x_1 (\tau) 
- u(\tau)\right] d\tau ; \quad \text{ with } I(x_1^0, x_2^0) = x_2^0 + \beta x_1^0. \]
Then \begin{itemize} \item[(i)] for $u(t) = -1,$  
\begin{eqnarray*} \tilde{\Delta}_2(t) & =& 
I(\tilde{x}_{1.2}, \tilde{x}_{2.2}) - I(x_1^0, x_2^0)
= t(-\sin x_1^0-1) - \frac{t^2}{2} x_2^0 \cos x_1^0 \\ 
&&+ \int_0^t \left\{ \sin \left[ x_1^0 + \tau x_2^0 + 
\frac{\tau^2}{2}(-\beta x_2^0 - \sin x_1^0 - 1) \right]+1 \right\} d\tau ; \\
\left|\tilde{\Delta}_2(t)\right| & \leq & 
\dfrac{t^2}{2} x_2^0  + 4 t \end{eqnarray*}
\[\text{leading to }  \left|\tilde{\Delta}_2(t)\right| < \kappa  \text{ for } 
t \in \left[0, \frac{-4 + \sqrt{16+ 2 x_2^0\kappa}}{x_2^0} \right], \text{ with }
\kappa \text{ a sufficiently small positive number}. \]
\item[(ii)] for $u(t) = 1,$ 
\begin{eqnarray*} \hat{\Delta}_2(t) & =& 
I(\hat{x}_{1.2}, \hat{x}_{2.2}) - I(x_1^0, x_2^0) = 
t(-\sin x_1^0+1) - \frac{t^2}{2} x_2^0 \cos x_1^0 \\ 
&&+ \int_0^t \left\{ \sin \left[ x_1^0 + \tau x_2^0 + 
\frac{\tau^2}{2}(-\beta x_2^0 - \sin x_1^0 + 1)\right]-1\right\} d\tau ;\\
\left|\hat{\Delta}_2(t)\right| & \leq & \dfrac{t^2}{2} x_2^0 + 4 t,\end{eqnarray*}
\[\text{ yielding }  \left|\hat{\Delta}_2(t)\right| < \kappa \text{ for }
t \in \left[0, \frac{-4 + \sqrt{16+ 2 x_2^0\kappa}}{x_2^0} \right], \text{ with } \kappa 
\text{ a sufficiently small positive number}. \]  
\end{itemize}
 
\subsection{Problem 3}\label{sec:5}
Let us examine the optimal control problem when the 
state equation is the Van der Pol equation \cite{Arnold,Conti}
\begin{equation}\ddot{x} + \varepsilon\dot{x}(x^2 - 1) + x = u(t) \end{equation}
such that the control $u(t)\in \left[\alpha,\beta\right].$ 
The problem has the form

\begin{equation} \label{prob301} \left\{ \begin{array}{lcl} 
\dot{x}_1 &=& x_2 \\ \dot{x}_2 &=& - x_1 + 
\varepsilon x_2 (1 - x_1^2) + u(t) \\ 
x_1(0) &=& x_1^0 ; \quad  x_2(0) = x_2^0 ; 
\\x_1(T) &=& 0  ; \quad x_2(T) = 0 ;  \\ 
\alpha &\leq& u \leq \beta \\ T & \longrightarrow & 
\inf \end{array} \right. \end{equation}
where $\varepsilon, \alpha, \beta$ are real constants.

The Hamilton-Pontryagin's function is 
\begin{equation} \eta (x, u, \psi) =  -1 + \psi_1 x_2 + \psi_2 
\left[- x_1 + \varepsilon x_2 (1 - x_1^2) 
+ u(t) \right].  \end{equation}
According to the Pontryagin's maximum 
principle, the supremum of the function $\eta$ 
depending on $x_1, x_2, \psi_1, \psi_2, u$ with respect to $u$
is reached when the control takes the following form:
\[ u(t) = \left\{\begin{array}{lc}  
\beta & \text{  if  }  \psi_2(t) > 0 \\
\alpha & \text{  if  }  \psi_2(t) < 0  \end{array}\right. \] 
Two cases are examined 
\begin{itemize} \item[(i)] Taking $u(t)= 
\alpha,$ we get the following Cauchy problem 
\begin{equation}\label{prb305}\left\{\begin{array}{l} \dot{x}_1 = x_2 \\ 
\dot{x}_2 = - x_1 + \varepsilon x_2 (1 - x_1^2) + \alpha \\ 
\dot{\psi}_1 = (1 + 2 \varepsilon x_1 x_2) \psi_2 \\ 
\dot{\psi}_2 = - \left[\psi_1 + \varepsilon (1 - x_1^2) \psi_2\right] \\       
x_1(0) = x_1^0 ; \quad x_2(0) = x_2^0 ; \quad 
\psi_1(0) = \psi_1^0 ; \quad \psi_2(0) = \psi_2^0,
\end{array}\right.\end{equation}

which, according to (\ref{SODE1}), can be reduced to 
\begin{equation}\label{prb309}\left\{\begin{array}{l} \dot{y}_1 = y_2 \\ 
\dot{y}_2 = - y_1 + \varepsilon y_2 (1 - y_1^2) + \alpha \\ 
\dot{y}_3 = (1 + 2 \varepsilon y_1 y_2) y_4 \\ 
\dot{y}_4 = - \left[y_3 + \varepsilon (1 - y_1^2) y_4 \right] \\         
y_1(0) = y_1^0 ; \quad y_2(0) = y_2^0 ; \quad 
y_3(0) = y_3^0 ; \quad y_4(0) = y_4^0. \end{array}\right.\end{equation}

The Hamiltonian of the system is computed as follows 
\begin{eqnarray}\label{prb313} H  = p_1 y_2^0 + p_2 \left\{- y_1^0 
+ \varepsilon y_2^0 \left[1 - (y_1^0)^2\right] + \alpha \right\} 
+ p_3 (1 + 2 \varepsilon y_1^0 y_2^0) y_4^0 \nonumber\\+ p_4 \left\{ y_3^0 
+ \varepsilon \left[1 - (y_1^0)^2\right] y_4^0\right\}+ i\varepsilon \left[1 - (y_1^0)^2\right]  . \end{eqnarray}
The first order approximation solutions are given by
\begin{eqnarray*} \begin{pmatrix} \left[ H,  y_1^0 \right] \\ \left[H,  y_2^0 \right] \\
\left[ H, y_3^0 \right]\\ \left[H,  y_4^0 \right] \end{pmatrix} & =& 
\begin{pmatrix} -i  y_2^0 \\ -i \left\{-y_1^0 + 
\varepsilon y_2^0  \left[1 - (y_1^0)^2\right] +\alpha \right\}\\ 
-i \left[( 1 + 2 \varepsilon y_1^0 y_2^0 )y_4^0 \right]\\
-i \left\{ y_3^0 + \varepsilon \left[1 - (y_1^0 )^2\right] y_4^0\right\} \end{pmatrix} ; \\
\begin{pmatrix} \tilde{y}_{1.1} (t) \\ \tilde{y}_{2.1} (t) \end{pmatrix} & = & 
\begin{pmatrix} y_1^0 + t y_2^0 \\ y_2^0 + t \left\{-y_1^0 + 
\varepsilon y_2^0  
\left[1 -(y_1^0)^2\right]+\alpha  \right\} \end{pmatrix}. 	\end{eqnarray*}
In the original variables \[ \begin{pmatrix} \tilde{x}_{1.1} (t) \\  \tilde{x}_{2.1} (t) \end{pmatrix} 
= \begin{pmatrix} x_1^0 + t x_2^0 \\ x_2^0 + t 
\left\{-x_1^0 + \varepsilon x_2^0  \left[1 -(x_1^0)^2\right]+\alpha \right\}
\end{pmatrix}.\]

\item[(ii)]  For $u(t)= \beta,$ the corresponding derived Cauchy problem is put in the form:
\begin{equation}\label{prb329}\left\{\begin{array}{l} \dot{x}_1 = x_2 \\ 
\dot{x}_2 = - x_1 + \varepsilon x_2 (1 - x_1^2) + \beta \\ 
\dot{\psi}_1 = (1 + 2 \varepsilon x_1 x_2) \psi_2 \\ 
\dot{\psi}_2 = - \left[\psi_1 + \varepsilon(1 - x_1^2) \psi_2\right] \\
x_1(0) = x_1^0 ; \quad x_2(0) = x_2^0 ; \quad 
\psi_1(0) = \psi_1^0 ; \quad \psi_2(0) = \psi_2^0. \end{array}\right.\end{equation}
This SODE differs from (\ref{prb309}) only by the term $\beta$ replacing $\alpha$. 
Then, replacing {\it mutatis mutandis} $\alpha$ by $\beta,$ the SODE remains the same as in 
(\ref{prb309}). Hence, the first order approximation solutions are given in the form:
\[ \begin{pmatrix}	\hat{y}_{1.1} (t) \\ \hat{y}_{2.1} (t) \end{pmatrix} = 
\begin{pmatrix} y_1^0 + t y_2^0 \\y_2^0 + t \left\{-y_1^0 
+ \varepsilon y_2^0  \left[1 -(y_1^0)^2\right]+ \beta  \right\} 
\end{pmatrix} \] or, equivalently, in terms of the original variables  
\[ \begin{pmatrix}\hat{x}_{1.1} (t)\\\hat{x}_{2.1} (t)  \end{pmatrix}	 = 
\begin{pmatrix}	x_1^0 + t x_2^0\\x_2^0 + t \left\{-x_1^0 
+ \varepsilon x_2^0  \left[1 -(x_1^0)^2\right]+ \beta  \right\}
\end{pmatrix}.\]	

Taking into account the relations 
\[ \left\{\begin{array}{l} \tilde{x}_{1.1} (T) = 0 \\ 
\tilde{x}_{2.1}  (T) = 0 \end{array}\right. 
\text{ and  } \left\{\begin{array}{l} \hat{x}_{1.1} (T) = 0 \\ 
\hat{x}_{2.1}  (T) = 0 \end{array}\right.,\]
we get the following results
\[ \tilde{T} = \frac{x_2^0}{x_1^0 - \varepsilon x_2^0 
\left[1 -(x_1^0)^2\right]-\alpha}>0 \qquad \textrm{ and  } \qquad
\hat{T} =  \frac{x_2^0}{x_1^0 - \varepsilon x_2^0 \left[1 -(x_1^0)^2\right]-\beta} > 0, \]
with the relations \[  (x_2^0)^2 + (x_1^0)^2 - 
\varepsilon x_1^0 x_2^0 \left[1 -(x_1^0)^2\right]-\alpha  x_1^0 = 0\quad \text{    or    }\quad 
(x_2^0)^2 + (x_1^0)^2 - \varepsilon x_1^0 x_2^0 \left[1 -(x_1^0)^2\right]-\beta x_1^0 = 0.\] \end{itemize}
The optimal time solution to the problem is
\[ \min(\tilde{T}, \hat{T}) = \left\{\begin{array}{ll} \tilde{T} & 
\text{  if  } \dfrac{x_2^0}{x_1^0 - \varepsilon x_2^0 \left(1 -(x_1^0)^2\right)-\alpha} <
\dfrac{x_2^0}{x_1^0 - \varepsilon x_2^0 \left(1 -(x_1^0)^2\right)-\beta}, \\
\hat{T} & \text{ otherwise.} \end{array}\right.\]  
An integral of motion of the SODE (\ref{prob301}) 
is \begin{eqnarray}\label{prb343} 
J(x_1, x_2) & = &  x_2 + \varepsilon \left(\frac{1}{3} x_1^3 - x_1\right) + 
	\int_0^t \left[x_1(\tau) - u(t)\right] d\tau \\
J(x_1^0, x_2^0)	& = &  x_2^0 + \varepsilon \left(\frac{1}{3} 
	(x_1^0)^3 - x_1^0\right). \nonumber \end{eqnarray}
Setting  
\begin{itemize} \item[(i)] for $u(t) = \alpha,$ \begin{eqnarray*} \tilde{\Delta}_1(t) 
		& =& J(\tilde{x}_{1.1}, \tilde{x}_{2.1}) - J(x_1^0, x_2^0) = \frac{1}{3}\varepsilon(x_2^0)^3 t^3 + 
		\left[\frac{1}{2} x_2^0 + \varepsilon x_1^0 (x_2^0)^2\right]  t^2  \end{eqnarray*} 
and using the Cardan's formulas for cubic polynomials, we obtain 
\[ \left|\tilde{\Delta}_1(t)\right| < \kappa \quad \text{ for } 
t \in \left[0, \sqrt[3]{-\dfrac{q}{2} + \sqrt{Q}}+ 
\sqrt[3]{-\dfrac{q}{2} - \sqrt{Q}} - \dfrac{A}{3} \right[ \]
 with $\kappa$ a sufficiently small positive number and
\begin{eqnarray*}	Q & = & \left(\dfrac{p}{3}\right)^3 +\left(\frac{q}{2}\right)^2, 
\quad p = -\frac{1}{3} A^2, \quad q = \frac{2}{27} A^3 - B, \\
A & = & \frac{\frac{3}{2} + 3\left|\varepsilon x_1^0 x_2^0\right|}{\left|\varepsilon||x_2^0\right|^2}, \quad
B =  \frac{3\kappa}{\left|\varepsilon||x_2^0\right|^3} ;
\end{eqnarray*} 
\item[(ii)] and for $u(t) = \beta,$ 
\[ \hat{\Delta}_1(t) =\frac{1}{3}\varepsilon(x_2^0)^3 t^3 + \left[\frac{1}{2} x_2^0 + 
\varepsilon x_1^0 (x_2^0)^2\right]  t^2 ;	\]
we obtain $\left|\hat{\Delta}_1(t)\right| < \kappa$ 
for $t \in \left[0, \sqrt[3]{-\dfrac{q_1}{2} + \sqrt{Q_1}} 
+ \sqrt[3]{-\dfrac{q_1}{2} - \sqrt{Q_1}} - \dfrac{A_1}{3}\right[ $  

with $\kappa$ a sufficiently small positive number and
\begin{eqnarray*}	Q_1 & = &\left(\frac{p_1}{3}\right)^3 +\left(\frac{q_1}{2}\right)^2, 
\quad p_1 = -\frac{1}{3} A_1^2 , 
\quad q_1 = \frac{2}{27} A_1^3 - B_1 \\
	A_1 & = & \frac{\dfrac{3}{2} + 3\left|\varepsilon x_1^0 x_2^0\right|}{\left|\varepsilon||x_2^0\right|^2}, \quad
B_1 = \frac{3\kappa}{\left|\varepsilon||x_2^0\right|^3}.
\end{eqnarray*} 
\end{itemize}

\section{Concluding remarks}\label{sec:6}
In this paper, we have successfully extended the domain of applicability of Bagarello's approach, 
developed for the analysis of ordinary differential equations, to 
optimal control  problems.
The relevance of the derived approximation scheme (\ref{approx}) 
 and  error estimation  given in formula 
(\ref{approxError}) has been exploited to investigate 
time-optimal problems of forced harmonic oscillator systems.
Three particular cases have been explicitly treated and discussed.

\vspace{0.5cm}



\begin{thebibliography}{99}

\bibitem{Adomian} Adomian, G.: A review of the decomposition 
					method and some recent results for nonlinear equations. 
					Comput. Math. with Applic., \textbf{21}(5), 101-127 (1991).

\bibitem{Adomian1} Adomian, G.: The closure approximation in the hierarchy equations.
				Int. Jour. of Stat. Phys., \textbf{3}(2), 127-133 (1971).

\bibitem{Adomian2} Adomian, G.: Nonlinear stochastic systems theory 
and applications to physics. \textbf{46}, Mathematics and Its Applications, 
(Kluwer), Springer, (1989). 

\bibitem{Salamon} Andresen, B., Hoffmann, K. H., Nulton, J., Tsirlin, A., Salamon, P.:
Optimal control of the parametric oscillator.  Eur. J. Phys. \textbf{32}, 827-843 (2011).

\bibitem{Arnold} Arnold, V. I.: Equations diff\'erentielles. Mir, Moscou (1981).

\bibitem{Arutyunov} Arutyunov, A. V., Tynyanskii, N. T.: First-and second-order 
conditions in the problem of optimal high-speed. Uspekhi Mat. Nauk. 
\textbf{36}:6(222), 199-200 (1981);
English transl. in Russian Math. Surveys \textbf{36} (1981).

\bibitem{Bagarello97} Bagarello, F.: A Non-Commutative approach to ordinary differential equations. 
				Int. Jour. of Theoretical Physics, \textbf{43}(12), 2371-2394 (2004).

\bibitem{Cherruault} Cherruault, Y., Adomian, G.: Decomposition methods : a new proof of convergence. 
				Math. Comp. Modelling, \textbf{18}(12), 103-106 (1993).

\bibitem{Cherruault1} Cherruault, Y.: Convergence of Adomian's method Kybernetes. 
\textbf{18}(2), 31-38 (1989).

\bibitem{Conti} Conti, R.: Control and the Van der Pol equation. In J. F\'abera (ed) 
Equadiff IV, pp. 73-80, S. Lecture Notes in Mathematics, \textbf{703}, 
Springer Berlin-Heidelberg, (1979).
				
\bibitem{Dasanayake} Dasanayake, I.S.: Optimal control of weakly forced 
nonlinear oscillators (2013). All theses and dissertations (ETDs). paper 1084.
http://openscholarship.wustl.edu/etd/1084

\bibitem{Barone} Flores-Hidalgo, G., Barone, F. A.: The one dimensional 
damped forced harmonic oscillator revisited. Eur. J. Phys. \textbf{32}, 377-388 (2011).

\bibitem{Gamkrelidze} Gamkrelidze, R.V.: Principles of optimal control theory.
 					2nd ed., Izdat. Tbilis. Univ., Tbilisi (1977) ; English transl., Plenum Press, New York (1978).

\bibitem{Holmes} Holmes, M. H.: Introduction to perturbation methods. 2d edition, Springer, New York (2012).

\bibitem{Hounkonnou} Hounkonnou, M. N.,  Dkengne Sielenou, P. A.: On factorizable classes 
				of second order linear differential equations with rational functions coefficients. 
				SUT Jour. of Math., \textbf{46} (2), 2015-229 (2010).

\bibitem{Sielenou} Hounkonnou, M. N., Dkengne Sielenou, P. A.: Adomian method for 
underdetermined systems of differential equations.
Afr. Diaspora J. Math. (N.S.) \textbf{12} (2), 73-103 (2011).

\bibitem{Krasnov} Krasnov, M., Kiss\'elev, A., Makarenko, G. :
				Recueil de probl\`emes sur les \'equations diff\'erentielles ordinaires. 
				Mir, Moscou (1978) (trad. fr. 1981).

\bibitem{LandauI} Landau, L., Lifchitz, E.: Physique th\'eorique, 
Tome 1, M\'ecanique. Mir, Moscou (1982).

\bibitem{MoshinskyAl} Moshinsky, M., and Smirnov, Y. F.: 
The harmonic oscillator in modern physics. Informa HealthCare, Amsterdam (1996).

\bibitem {Nayfeh} Nayfeh, Ali H.: Perturbation methods. Wiley, New York (2000).

\bibitem{PontryaginAl} Pontryagin, L.S. et al.: The mathematical theory of optimal processes.
 2nd ed., Nauka, Moscow (1969) ; English transls. of 1st ed., Wiley (1962) and Macmillan (1964).

\bibitem{ReedSimon I} Reed, M., Simon, B.: Methods of Modern 
Mathematical Physics, I. Academic Press, New York (1980). 

\bibitem{Stetter} Stetter, Hans J.: Analysis of discretization methods for 
ordinary differential equations. Springer, Berlin-Heidelberg (1973).

\bibitem{Tynyanskii} Tynyanskii, N. T., Arutyunov, A. V.: On the system of Jacobi 
equations in a time-optimal problem. Izv. Akad. 
Nauk SSSR Ser. Mat. \textbf{46}: 5, 1082-1105 (1982);
English transl. in Math. USSR Izv. \textbf{21} (1983).

\bibitem{VanDooren} Van Dooren, R.: Numerical study of the 
controlled Van der Pol oscillator in Chebyshev series. 
Journ. of Appl. Math. and Phys. (ZAMP). \textbf{38}(6), 934-939 (1987).

\bibitem{Vassiliev} Vassiliev, F. L. V.: Numerical methods for 
the optimization problems. Nauk, Moscou (1988) (in Russian).

\end{thebibliography}
\end{document}